\documentclass[prb,aps,twocolumn]{revtex4}
\usepackage{amsmath}
\usepackage{amssymb}
\usepackage{graphicx}
\def\sigmaxc{\stackrel{\leftrightarrow}{\sigma}}

\begin{document}
\title{Electronic viscosity in a quantum well: A test for the local density approximation}

\author{Roberto D'Agosta}
\email{dagosta@physics.ucsd.edu}
\affiliation{University of California - San Diego, La Jolla, CA 92093}

\author{Massimiliano Di Ventra}
\affiliation{University of California - San Diego, La Jolla, CA 92093}
\author{Giovanni Vignale}
\affiliation{University of Missouri - Columbia, Columbia, MO 65211}

\begin{abstract}
In the local density approximation (LDA) for electronic
time-dependent current-density functional theory (TDCDFT) many-body
effects are described in terms of the visco-elastic constants of the
homogeneous three-dimensional electron gas.  In this paper we
critically examine the applicability of the three-dimensional LDA to
the calculation of the viscous damping of 1-dimensional collective
oscillations of angular frequency $\omega$  in a quasi 2-dimensional
quantum well.  We calculate the effective viscosity $\zeta(\omega)$
from perturbation theory in the screened Coulomb interaction and
compare it with the commonly used three-dimensional LDA  viscosity
$Y(\omega)$. Significant differences are found.   At low frequency
$Y(\omega)$ is dominated by a shear term, which is absent in
$\zeta(\omega)$.  At high frequency $\zeta(\omega)$ and $Y(\omega)$
exhibit different power law behaviors ($\omega^{-3}$ and
$\omega^{-5/2}$ respectively), reflecting different spectral
densities of electron-hole excitations in two and three dimensions.
These findings demonstrate the need for better approximations for
the exchange-correlation stress tensor in specific systems where the
use of the three-dimensional functionals may lead to unphysical
results.
\end{abstract}
\date{\today}
\maketitle

\section{Introduction}
Time-dependent density functional theory is  one of the premier
techniques for  the study of the dynamics of quantum many-body
systems.\cite{Runge1984,Marques,Giulianivignale} The central idea of the theory,
dating back to the pioneering work of Hohenberg, Kohn and
Sham,\cite{Hohenberg1964,Kohn1965} is that 
the interacting many-body system can be
simulated  by a non-interacting system that yields the same density
under the action of the self-consistent electrostatic potential
$V_H$  and an additional exchange-correlation (xc) potential
$V_{xc}$ that is uniquely determined by the particle density $n$.\cite{vanLeeuwen1999}
This approach, while rigorous in principle, runs into serious
difficulties when applied to phenomena that involve dissipation and
memory -- for example, the  damping of collective modes.  This is
because $V_{xc}$ is a strongly non-local functional of the
density.\cite{Vignale1996}

About 20 years ago a time-dependent current density functional theory
(TDCDFT) was proposed,\cite{Ghosh1988,Vignale1996} which offered a natural way
to treat memory and dissipation without losing the advantages of a
local description of many-body effects.  In this theory the
exchange-correlation potential was replaced by an exchange-correlation
force field $\vec F_{xc}=-e \vec E_{xc}$  (the notation is designed
for electronic systems:  $-e$ is the electron charge and $\vec E_{xc}$
is an ``exchange-correlation electric field") which was represented as
the divergence of a stress tensor $\sigmaxc$
\begin{equation}
\label{exc}
\vec F_{xc} = -e \vec E_{xc}= \frac{1}{n}\vec \nabla \cdot \sigmaxc~,
\end{equation}
which in turn was expressed as a local functional of the equilibrium
density $n$ and the velocity $\vec v = \frac{\vec j}{n}$, where
$\vec j$ is the particle current density. 

In the linear response regime an approximate form of the stress
tensor was proposed,\cite{Vignale1996,Vignale1997b} which has the same form as
the stress tensor of the classical Navier-Stokes
hydrodynamics:~\cite{DAgosta2006a}
\begin{equation}\label{sigmaxc}
\sigma_{xc,ij}=-p_{xc}\delta_{ij} + \eta \left(\frac{\partial
    v_i}{\partial r_j}+\frac{\partial v_j}{\partial r_i} - \frac{2}{3}
  \vec \nabla \cdot \vec v \delta_{ij}\right)+\zeta  \vec \nabla \cdot
\vec v \delta_{ij}~,
\end{equation}
where $p_{xc}$ is the exchange-correlation pressure of the homogeneous
electron gas at the local density $n$,  $\eta$ is the {\it shear
  viscosity} and  $\zeta$ is the {\it bulk viscosity} of the
homogeneous electron gas at the local equilibrium density
$n_0$. $\eta$ and $\zeta$ are frequency-dependent and have imaginary
parts which are related to a shear modulus and a dynamical bulk
modulus respectively.

An essential feature of this approximation is that it is still local
in the equilibrium density and the velocity field.  In particular, the
adiabatic-local density approximation (3D-LDA) amounts to keeping only
the first term on the rhs of Eq.~(\ref{sigmaxc}).

The exchange-correlation field~(\ref{exc}) with stress
tensor~(\ref{sigmaxc})  has been applied to the study of several
systems (semiconductor quantum wells,
\cite{Williams2001,Ullrich2001,Ullrich2002} 
atoms,\cite{Ullrich2004} semiconductors and polymers,
\cite{vanFaassen2002,Berger2005,Berger2007} 
molecular junctions\cite{Sai2005,DAgosta2006a,DAgosta2006c,Sai2007}
and metals\cite{Romaniello2005,Berger2006}) with
varying degrees of success.  However a fundamental difficulty exists.
Because the viscosities $\eta$ and $\zeta$ are borrowed from an
infinite three-dimensional electron gas, this approximation
implicitly assumes the existence of a continuum spectrum of
excitations (electron-hole pairs).   Needless to say, this assumptions
is not justified in systems with discrete energy levels.  This may
lead in some cases to spurious results.  For example, the optical
transitions between discrete energy levels of atoms are found to have
a spurious linewidth.\cite{Ullrich2004}
\begin{figure}[t]
\includegraphics[width=8cm,clip]{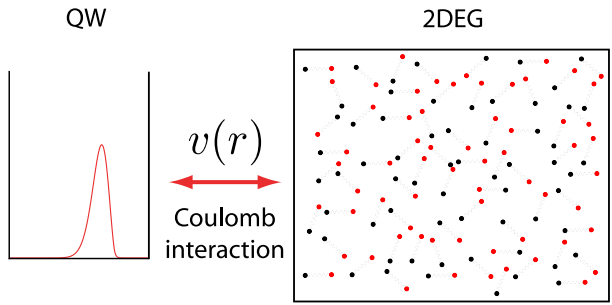}
\caption{Two-subsystems model for electrons in a two-dimensional
  quantum well. The collective motion of the electrons in the $z$
  direction is coupled, through the Coulomb interaction, to density
  fluctuations in the plane of the 2DEG. Energy is exchanged between
  the two subsystems via the creation of electron-hole pairs in the
  2DEG.}
\label{twosystems}
\end{figure}

A more ``friendly" system is the electron gas in a two-dimensional
quantum well (shown in Fig.~\ref{twosystems}), in which the
electrons perform collective oscillations  in a direction
perpendicular to the plane of the quantum well (the $z$-direction),
while remaining homogeneous in the plane of the well (the $x-y$
plane).   Spontaneous oscillations of this kind -- known as {\it
intersubband plasmons} -- have been intensively studied because of
their possible use in devices in the terahertz frequency range.  In
practice these oscillations, spontaneous or forced,  provide one of
the best testing grounds for TDDFT calculations of damping and
relaxation.\cite{DAgosta2006,Wijewardane2005,Ullrich2004}  Even in
this favorable case, however,  one may question the accuracy of the
local density approximation for the viscosity.  To begin with, the
electron-hole pairs that are responsible for the damping and the
screening of the electron-electron interaction should be those of an
essentially two-dimensional electron gas (2DEG) --  not those of a
three-dimensional electron gas as implicitly assumed in the 3D-LDA.
However, the system is not exactly  two-dimensional, even if one
neglects inter-subband transitions,  because the electron-electron
interaction in the lowest subband is modified by {\it form factors},
which take into account the finite extent of the wave function in
the $z$ direction.  More importantly, in a collective oscillation
that preserves uniformity  in the $x-y$ plane, the Fermi surface of
the 2DEG does not change its (circular) {\it shape} (see
Fig.~\ref{fermisurface}): this means that  only the {\it bulk
viscosity}, $\zeta$, contributes to the damping of the collective
mode.  By contrast the dominant contribution to the damping in
3D-LDA  comes from the {\it shear viscosity}, which is associated
with changes in {\em shape} of the implicitly assumed
three-dimensional Fermi surface.  Finally, it is evident that the
above difficulties cannot be solved by resorting to a strictly
two-dimensional LDA, since the system becomes homogeneous and
time-independent when strictly projected in the $x-y$ plane.
\begin{widetext}
\begin{center}
\begin{figure}[h!] 
\includegraphics[width=13cm,clip]{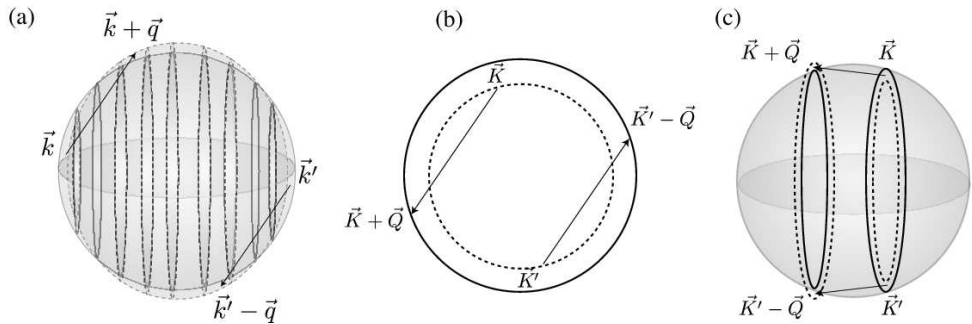}
\caption{(a) In a three-dimensional collective mode the Fermi surface changes shape periodically from oblate to prolate.  The process can be pictured as a transfer of electrons
between the Fermi disks into which the Fermi surface can be sliced, perpendicular to the direction of propagation of the mode.    These excitations are responsible for the finite shear viscosity of the electron liquid.   (b) In 2D, for an isotropic single band system the Fermi surface is a circle: excitations can only change the radius of the circle, not its shape.  There is no shear viscosity.  (c) In a multi-band isotropic  2D systems, inter-band transitions allow a change in shape of the Fermi surface -- formed in this case by several  parallel disks:  the shear viscosity reappears. (2D wave vectors are represented by capital letters).}
\label{fermisurface}
\end{figure}
\end{center}
\end{widetext}

In view of the above difficulties it is clearly of great interest to test
the validity of the 3D-LDA against a more direct calculation of the
damping rate of a collective oscillation -- a calculation that
explicitly takes into account the reduced dimensionality of the
quantum well.   Our idea is to study a particularly simple
oscillation that is described, to zero-order in the Coulomb
interaction, by a separable wave function, i.e. the product of an
assigned time-dependent function of the $z$ coordinate of the
electrons times a function of the $x$ and $y$ coordinates: the
latter may be either the ground state or a uniform excited state of
the quasi-2DEG in the $x-y$ plane.  For example,  we can put all the
electrons  in a given   time-dependent combination of the lowest and
the first excited subband:  we assume that there is an external
time-dependent potential that can do this. The main physical
assumption here is that all the electrons populate a {\em single}
time-dependent subband, and no inter-subband scattering is
considered. The time-dependent electronic density $n(z,t)$ creates,
via time-dependent form factors, a time-dependent Coulomb field
which induces transitions between different states of the 2DEG.  We
can make use of the Fermi Golden Rule to calculate the rate at which
the motion in the $z$-direction creates electron-hole pairs in the
$x-y$ plane. From this we calculate the rate at which energy is
transferred from the oscillatory motion in the $z$ direction to
electron-hole pair excitations in the $x-y$ plane.   In brief, the
electron-hole pairs in the $x-y$ plane act as a thermal bath for the
motion in the $z$ direction. Obviously this approach is justified in
the limit  of weak Coulomb interaction (high density).

It turns out that it is possible to recast the results for the damping
rate  in a form that is analogous to the LDA, except that it involves
a {\it non-local viscosity} $\zeta(z,z')$.  This non-local viscosity
can be converted into an  effective local viscosity $\zeta(z)$ by
integrating over  $z'$ at fixed $z$ -- a procedure that amounts to
neglecting  the finite range of nonlocality on the scale of variation
of the density.    In this manner we are able to compare the
approximate LDA viscosity $Y(z)$  -- a function of the local
equilibrium density --  with the more accurate one, $\zeta(z)$.

We find significant differences  between $\zeta(z,\omega)$ and
$Y(z,\omega)$.  At low frequency the latter is dominated by a shear
term, which is absent in $\zeta$.  For this reason $\zeta$ turns out
to be numerically much smaller than $Y$.  At high frequency
$\zeta(z,\omega)$ and $Y(z,\omega)$ exhibit different power law
behaviors ($\omega^{-3}$ and $\omega^{-5/2}$ respectively),
reflecting different spectral densities of electron-hole excitations
in two and three dimensions.

We evaluate the damping rate as a function of the frequency of the
external field, $\omega$, for the screened and unscreened
potentials. We find, as one could expect, that the rate of energy
dissipation in the screened (unscreened) interaction goes as
$\omega^4$ ($\omega^2$) at small frequencies. 
In the 3D-LDA
approximation, instead, 
the energy dissipation
rate goes as $\omega^2$. Thus the 3D-LDA largely overestimate the energy
dissipation.

It is already known\cite{Ullrich2004} that the LDA viscosity is
spurious in small systems like atoms and molecules.  The present
findings show that a serious loss of accuracy can occur also in
infinite systems,  when the motion is restricted to a single
one-dimensional subband as in the example considered here.
Therefore, one needs to apply special care when using 3D-LDA
approximations in specific systems such as the one we consider here.
The development of a better approximation for the viscous stress
tensor -- an approximation that is uniformly applicable across
different dimensionalities  -- thus emerges as a critical issue.

\section{The model}
The Hamiltonian for an interacting N-electron system confined to a
quantum well (QW) of width $L$  in the $z$ direction is
\begin{equation}
\hat H=\sum_i \frac{\hat p_i^2}{2m}+V(\hat
r_i)+\frac12\sum_{i\not=j}v(|\hat r_i-\hat r_j|)
\end{equation}
where $v(r)=e^2/4\pi \epsilon |r|$ is the Coulomb potential ($\epsilon$ is the background
dielectric constant) and
\begin{equation}
V(r)\equiv V(z)=\left\{
\begin{array}{cc}
0 & |z|<L/2\\
\infty & |z|>L/2
\end{array}
\right. .
\end{equation}

In the absence of electron-electron interactions ($v(r)\equiv 0$) we
can separate the Hamiltonian into two parts describing respectively
the motion in the $z$ direction and the motion of the
two-dimensional electron gas (2DEG) in the plane:
\begin{equation}
\hat H_0=\hat H_z+\hat H_{2DEG}
\end{equation}
where
\begin{eqnarray}
&&\hat H_z=\sum_{i=1}^N \frac{p_{z,i}^2}{2m}+V(\hat z_i)+V_{ext}(\hat z,t),\\
&&\hat H_{2DEG}=\sum_{i=1}^N
\frac{p_{x,i}^2+p_{y,i}^2}{2m}\equiv\sum_{i=1}^N \frac{P_i^2}{2m},
\end{eqnarray}
where $V_{ext}(z,t)$ is the external potential that excites the QW (see the following discussion).
Obviously the Coulomb interaction couples these two subsystems in a
non-trivial way. In the following we will consider the Coulomb
interaction as a small perturbation and then apply time-dependent
perturbation theory to study the energy exchange between the
$z$-motion and the density fluctuations of the 2DEG.  In particular,
we will show that, if one assumes a given motion in the $z$ direction,
the Coulomb interaction excites electron-hole pairs in the $x-y$
plane, thus effectively transferring energy from the $z$-motion to the
2DEG.

A solution to the Schr\"odinger equation with the unperturbed
Hamiltonian $\hat H_0$ can be written as the product of solutions of
$H_{2DEG}$ and $H_z$ in the following manner:
\begin{equation}
|\Psi\rangle=|\Phi\rangle_{2DEG}|\chi\rangle_z
\label{state}
\end{equation}
where $|\Phi\rangle_{2DEG}$ is a Slater determinant of $N$ plane waves in the coordinates $R_i$  and
\begin{equation}
|\chi\rangle_z= |\chi\rangle_1 \times   |\chi\rangle_2 ... \times |\chi\rangle_N
\label{quantumwell}
\end{equation}
is a symmetric function of the $z$ coordinates of the electrons.   We
assume   that all the electrons share a common motion in the $z$
direction  specified by the one-electron wave function $\chi(z,t)$,
which is  sustained by a suitable time-dependent external potential
$V_{ext}(z,t)$.  The density remains uniform in the $x-y$ plane. The
antisymmetry of the complete wave function is ensured by the
antisymmetry of  $|\Phi\rangle_{2DEG}$.

Next we turn on the electron-electron interaction, and derive an
effective hamiltonian for the dynamics of  $|\Phi\rangle_{2DEG}$ in
the presence of the driving potential $V_{ext}(z,t)$.   This is done
by substituting Eqs. (\ref{state}) and (\ref{quantumwell}) into the
Sch\"odinger equation
\begin{equation}
i\hbar \partial_t |\Psi(t)\rangle=\hat H(t) |\Psi(t)\rangle.
\end{equation}
Making use of the Fourier representation of the Coulomb potential,
\begin{equation}
v(\hat r)=\frac{e^2}{\epsilon V}\sum_{q,Q}\frac{e^{-i q \hat z}e^{-i Q\cdot R}}{q^2+Q^2}
\end{equation}
($R$ and $Q$ are vectors in the 2D real and momentum space
respectively, with $r=(R,z)$, $q$ is a 1D momentum and $V$ the
volume), and the Fourier transform of the density
\begin{equation}
n(q,t)=\int_{-L/2}^{L/2}dz~e^{iq z}n(z,t)
\end{equation}
we can write the electron-electron interaction as
\begin{equation}\label{HEE}
\hat H_{ee}(t) =\frac{e^2}{2\epsilon V}\sum_{i\not =j}\sum_{q,Q}\frac{e^{-iq  (\hat z_i- \hat z_j)}}{q^2+Q^2}e^{-iQ\cdot (\hat R_i-\hat R_j)}.
\end{equation}
Assuming that the total wavefunction is still approximately of the
form~(\ref{state}) we can ignore the correlation between the $z$'s and
the $R$'s, and simply replace $e^{-iq  (\hat z_i- \hat z_j)}$ by its
average value in the given one electron wave
function~(\ref{quantumwell}):
\begin{equation}
e^{-iq  (\hat z_i- \hat z_j)} \to \left\langle e^{-iq  (\hat z_i- \hat
    z_j)} \right\rangle = |n(q,t)|^2,
\end{equation}
independent of $i$ and $j$. Then the effective dynamics for
$|\Phi\rangle_{2DEG}$ is given by
\begin{equation}
i\hbar \partial_t |\Phi\rangle_{2DEG}=\hat H_{eff} |\Phi\rangle_{2DEG}
\end{equation}
where
\begin{equation}
\hat H_{eff}=\sum_{i=1}^{N}\frac{\hat P_i^2}{2m}+\frac{e^2}{2\epsilon
  V}\sum_{i\not =j}\sum_{q,Q}\frac{|n(q,t)|^2}{q^2+Q^2}e^{-iQ\cdot
  (\hat R_i-\hat R_j)}.
\end{equation}
Notice that $|n(q,t)|^2$ plays the role of a {\it time-dependent form
  factor} for the electron-electron interaction in the plane.
\begin{widetext}
To apply perturbation theory we assume that $n(q,t)$ is the sum of a
static part plus a small dynamical contribution:
\begin{equation}\label{nqt}
n(q,t)=n_0(q)+\delta n(q,t)
\end{equation}
with $|\delta n(q,t)|\ll n_0(q)$ for all $q$ and $t$. 
Neglecting  contributions of
order $(\delta n)^2$ we get the effective time-dependent Schr\"odinger
equation for $|\Phi\rangle_{2DEG}$
\begin{equation}
\left[i\hbar\partial_t -\sum_{i=1}^{N}\frac{\hat P_i^2}{2m}
-\frac{e^2}{2\epsilon V}\sum_{i\not
  =j}\sum_{q,Q}\frac{|n_0(q)|^2}{q^2+Q^2}e^{-iQ\cdot (\hat R_i-\hat
  R_j)}
-\frac{e^2}{\epsilon V}\sum_{i\not =j}\sum_{q,Q}\frac{Re[n_0(q)\delta
  n(-q,t)]}{q^2+Q^2}e^{-iQ\cdot (\hat R_i-\hat
  R_j)}\right]|\Phi\rangle_{2DEG}=0.
\label{effectivedynamics}
\end{equation}
\end{widetext}
The left hand side of Eq. (\ref{effectivedynamics}) is the sum of a
static part and a time-dependent part. The static part includes the
effect of the QW static form factor on the ground state and the
excited states of the 2DEG. We assume that the 2DEG is in its ground
state at $t=0$. The time-dependent part induces transitions between
eigenstates of the static 2DEG Hamiltonian and thus is responsible for
dissipation.

An important check for this model is that if the motion in the quantum
well is a rigid translation of the equilibrium density in the $z$
direction (Kohn mode), i.e.,  $n(z,t)=n_0(z+u(t)) \simeq n_0(z)
+ \partial_z n_0(z)u(t)$ where $u(t)$ is a {\it spatially uniform}
function of time, then the dissipation must be absent. This is easily
verified from Eq. (\ref{effectivedynamics}): for the Kohn mode we have
$\delta n(q,t)=iq n_0(q) u(t)$ where $u(t)$ is a real function, and
\begin{equation}
Re[n_0(q)\delta n(-q,t)]=Re[iq |n_0(q)|^2 u(t)]=0~.
\end{equation}
Therefore only the static non-dissipative part of the Hamiltonian
survives for the Kohn mode.

\section{Perturbation theory}
To apply the Fermi golden rule we assume that the forcing density
$\delta n(z,t)$ is a periodic function of time with angular frequency
$\omega$,  $\delta n(q,t) = \delta n(q,\omega)e^{-i\omega t}+ c.c.$,
and we write
\begin{equation}
\delta n(q,\omega) =  \frac{q j(q,\omega)}{\omega}
\end{equation}
where $j(q,\omega)$ is the Fourier amplitude of the current 
\begin{equation}
n_0(z)\partial_t u(z,t) = j(q,\omega) e^{i(qz-\omega t)} +~c.c.
\end{equation}
~~~~~~~~\\
The  energy transferred per unit of time by the $z$-motion to the 2DEG
is equal to the energy   $\hbar \omega$  absorbed in each allowed
transition  from the ground-state to an excited state of the 2DEG
times the rate at which the transition occurs.  Evidently the only
allowed transitions in leading order are the ones in which two
electron-hole pairs with opposite wave vectors $Q$ and $-Q$ are
created in the 2DEG, with energies adding up to $\hbar \omega$.  The
first pair is created by promoting an electron from state $K$ within
the two-dimensional Fermi surface of the 2DEG to state $K+Q$ outside
the Fermi surface.  Similarly, the second pair is created by promoting
an electron from $K'$ to  $K'-Q$, with $K'$ inside and $K'-Q$ outside
the Fermi surface [see Fig.~\ref{fermisurface}, panel b)].
So the energy absorbed per unit time  is
\begin{widetext}
\begin{eqnarray}\label{DEDT}
\frac{dE}{dt}&=&\frac{4\pi e^4}{\epsilon^2 \hbar {\cal
    A}}\sum_{K,K'<K_F}\int d^2Q \int dq\int dq'
\delta(\omega-\omega_{K,K',Q})\frac{q
  q'}{\omega}n_0(q)j(-q,\omega)\nonumber\\
&&\times
\frac{(1-n_{K+Q})n_K(1-n_{K'-Q})n_{K'}}{q^2+Q^2}\left[\frac{1}{q'^2+Q^2}-\frac{1}{q'^2+(Q+K-K')^2}\right]n_0^*(q')j^*(-q',\omega)
\end{eqnarray}
where $n_K$ is the Fermi occupation number [$n_K=\Theta(K_F-K)$, where
$\Theta(x)$ is the step function and $K_F$ is the Fermi wave vector]
and $\omega_{K,K',Q}\equiv \hbar [Q^2+Q\cdot(K'-K)]/m$ is the energy
of the double electron-hole pair. ${\cal A}$ is the area occupied by
the 2DEG.
The two terms in the square bracket arise from direct and exchange
processes respectively -- the latter arising from the antisymmetry of
the 2DEG wave function.  In the following we will disregard the
exchange term.  Aside from the fact that  its contribution is  small
in the high density limit, we must keep in mind that the exchange
contribution was also dropped in previous calculations of the LDA
viscosity:\cite{Conti1999}  for a fair comparison we must drop it here
too.

Introducing in Eq.~(\ref {DEDT}) the imaginary part of the
two-dimensional Lindhard function\cite{Giulianivignale}
\begin{equation}
 \Im m \chi_0^{2D}(Q,\omega) =-\frac{\pi}{\hbar{\cal A}}\sum_K
 n_K(1-n_{K+Q})\delta(\omega-\omega_{K,Q})~,
 \end{equation}
 where $\omega_{K,Q}=\hbar [Q^2/2m-Q\cdot K/m]$  we obtain
\begin{equation}
\frac{dE}{dt}=\frac{e^4\hbar {\cal A}}{4\pi^4 \epsilon^2 \omega}\int
dq~qj(-q,\omega)n_0(q)\int dq'~q'j^*(-q',\omega)n_0^*(q')\int d^2Q
\int_0^\omega d\omega'~\frac{\Im m\chi_0^{2D}(Q,\omega') \Im
  m\chi_0^{2D}(Q,\omega-\omega')]}{(q^2+Q^2)(q'^2+Q^2)}.
\end{equation}
Notice that $\chi_0^{2D}$ depends only on the sheet density of the
electron gas, which is independent of $z$.

As a last step we express this result in terms of the velocity field,
$v(z,t)=\partial_t u(z,t)=j(z,t)/n_0(z)$,
\begin{equation}
\frac{dE}{dt}=-2{\cal A}\int dz~\int dz'\partial_z
v(z,\omega)\zeta(z,z';\omega)\partial_{z'} v^*(z',\omega)
\label{energydissipation}
\end{equation}
where
\begin{equation}
\zeta(z,z';\omega)=n_0(z)n_0(z')\frac{\hbar}{4 \pi
  \omega}\int_{0}^\omega d\omega' \int dQ~Q \tilde v(Q,z)  \tilde
v^*(Q,z')\Im m\chi_0^{2D}(Q,\omega')  \Im
m\chi_0^{2D}(Q,\omega-\omega')
\label{bulkviscosity}
\end{equation}
is the non-local bulk viscosity of the electron gas in the QW and
\begin{equation}\label{vtilde}
\tilde v(Q,z)= \frac{2e^2}{\epsilon}\int_{-\infty}^\infty dq
\frac{n_0(q)e^{-iq z}}{q^2 + Q^2}
\end{equation}
the effective electron-electron interaction in the plane of the 2DEG.

Eq.~(\ref{energydissipation}) should be compared with the
corresponding LDA expression
\begin{equation}
\left.\frac{dE}{dt}\right\vert_{LDA}=-2{\cal A}\int dz |\partial_z
v(z,\omega)|^2 \left[\zeta^{LDA}(z;\omega)+\frac{4}{3}
  \eta^{LDA}(z,\omega)\right]~,
\label{energydissipation-LDA}
\end{equation}
where (see Refs.~\onlinecite{Conti1999} and \onlinecite{Qian2002})
\begin{eqnarray}
\label{3D-LDAViscosity1}
\zeta^{LDA}(z,\omega) &=& \frac{\hbar}{36\pi^2\omega}\int_0^\omega
d\omega' \int dq q^2 |v(q)|^2 \Im m\chi_0^{3D}(q,\omega') \Im  m
\chi_0^{3D}(q,\omega-\omega')\\
\eta^{LDA}(z,\omega) &=& \frac{4\hbar}{15 \pi^2 \omega}\int_0^\omega
d\omega' \int dq|v(q)|^2 q^2 \Im m\chi_0^{3D}(q,\omega') \Im
m\chi_0^{3D}(q,\omega-\omega')\nonumber\\
\label{3D-LDAViscosity2}
&&+\frac{\hbar}{5 \pi^2\omega^3}\int_0^\omega d\omega' \int dq
|v(q)|^2 q^4 \Im m\chi_{0T}^{3D}(q,\omega') \Im
m\chi_0^{3D}(q,\omega-\omega'),
\end{eqnarray}
$\chi_0^{3D}(q,\omega)$ is the 3D Lindhard function,
$\chi_{0T}^{3D}(q,\omega)$ is the noninteracting transverse
current-current response function in 3D, and $v(q)$ is the Fourier
transform of the screened 3D Coulomb interaction (about the screening
more will be said later).  The $z$-dependence of these functions
arises from the fact that the 3D Lindhard functions depend on the
local density.

To facilitate the comparison between the two sets of
formulas~(\ref{energydissipation})-(\ref{bulkviscosity})  and
~(\ref{energydissipation-LDA})-(\ref{3D-LDAViscosity2})  we
introduce the integrated viscosity
\begin{equation}
\zeta(z,\omega) \equiv \int \zeta(z,z';\omega)dz'~,
\end{equation}
which has the same physical dimensions as $\zeta^{LDA}(z,\omega)$
[Energy $\times$ Time/Volume] and can be directly compared to it.
The explicit expression for $\zeta(z,\omega)$ is
\begin{equation}
\zeta(z,\omega)=n_0(z)\frac{\hbar {\cal A}}{4 \pi
  \omega}\int_{0}^\omega d\omega' \int dQ~Q  \tilde v(Q,z)\tilde
v^*(Q) \Im m\chi_0^{2D}(Q,\omega')  \Im
m\chi_0^{2D}(Q,\omega-\omega')~,
\label{bulkviscosity-partialaveraged}
\end{equation}
where
$$
\tilde v(Q)\equiv \int  n_0(z)\tilde v(Q,z)dz~.
$$
\end{widetext}

It is interesting to compare  Eq.~
(\ref{bulkviscosity-partialaveraged}) with the expressions  obtained
from the xc kernel in the 3D-LDA [see Eq.~(\ref {3D-LDAViscosity1})].
In the homogeneous 3D electron liquid there are two independent
viscosities:  the bulk viscosity $\zeta(\omega)$ and the shear
viscosity $\eta(\omega)$.  The former arises from motions that
change the volume of the local Fermi surface (the density), but not
its shape. The latter appears when the motion changes the shape of
the Fermi surface, even if its volume does not change.  A
collisionless longitudinal collective mode, such as the ordinary
plasmon, involves both types of motion
simultaneously, as discussed in the caption of Fig.~\ref{fermisurface}:  the effective
viscosity for such a mode is
$Y(\omega)\equiv\zeta^{LDA}(\omega)+\frac{4}{3}\eta^{LDA}(\omega)$.\cite{Wijewardane2005,DAgosta2006} 
$Y$ can be easily constructed from the combination of the two
equations~(\ref {3D-LDAViscosity1}) and (\ref {3D-LDAViscosity2}).

Aside from the obvious difference in the dimensionality of the
Lindhard functions,  we see that our expression~(\ref{bulkviscosity})
for $\zeta(z,\omega)$  is formally similar to the
expression~(\ref{3D-LDAViscosity1}) for $\zeta^{LDA}(\omega)$ in
3D-LDA.  These expressions, however,  vanish as $\omega^2$ at low
frequency, because the Lindhard spectra $\Im m\chi_0^{2D
  (3D)}(q,\omega)$ vanish as $\omega$.  As a result $\zeta(z,\omega)$,
as well as $\zeta^{LDA}(\omega)$, goes as $\omega^2$ at low
frequency. On the other hand, the behavior of the LDA shear
viscosity, $\eta^{LDA}(\omega)$, is quite different.  From Eq.~(\ref
{3D-LDAViscosity2}) we see that this quantity contains a term
involving the transverse current spectrum $\Im m\chi_{0T}^{3D}$,
and this term tends to a finite limit for $\omega \to 0$  because
the factor $1/\omega^3$  compensates for the smallness of the
Lindhard spectra.

Therefore a fundamental difference exists between our results and
those of the 3D-LDA in the low frequency regime.  While the LDA
viscosity is dominated by the shear term, which remains finite for
$\omega \to 0$,  the present viscosity is purely of the bulk type and
vanishes as $\omega^2$ for $\omega \to 0$, at least when the screening
of the electron-electron interaction is properly taken into account
(see discussion below). Physically, the absence of a shear term is due
to the fact that the oscillatory motion in the $z$ direction does not
change the shape of the local Fermi surface of the 2DEG (a circle).
Whereas, in a 3D oscillation the local Fermi surface changes its shape
periodically from a prolate to an oblate ellipsoid (passing through
the sphere),  generating shear friction in the process (see
Fig.~\ref{fermisurface})

Before proceeding to a detailed comparison of the numerical results
for the viscosities and the energy dissipation rates we need to say
something more about the role of the screening of the
electron-electron interaction.  First of all, it must be noticed that
the integrals over wave vector in Eqs.~(\ref {3D-LDAViscosity1})-(\ref
{3D-LDAViscosity2}) diverge, in the limit $\omega \to 0$, if $v(q)$ is
taken to be the Fourier transform of the bare Coulomb interaction,
$4\pi e^2/\epsilon q^2$.  The divergence comes from the small
$q$ region.  The standard cure for this type of divergence is to
screen the interaction by the static dielectric function of the
electron gas, which, in the high-density limit, can be reasonably
approximated by the Thomas-Fermi formula. The screened interaction has
the form
\begin{equation}\label{3D-screened-interaction}
v(q) = \frac{4 \pi e^2}{\epsilon(q^2+\kappa^2)}~,
\end{equation}
where $\kappa = \sqrt{4 \pi e^2N(0)}$ is the Thomas-Fermi wave vector,
and $N(0)$ is the density of states at the Fermi energy.

The situation is somewhat different in Eq.~
(\ref{bulkviscosity-partialaveraged}).  Because the effective
interaction $\tilde v(Q,z)$ diverges  at small $Q$ only as $\frac{2
  \pi e^2}{Q}$,  it turns out that the wave vector integral is finite
even without including the screening. Nevertheless, the integral has a
large contribution from extremely small wave-vectors, of order
$\omega/v_F$ for $\omega \to 0$, which effectively changes the
formally expected $\omega^2$ behavior of  $\zeta(z,\omega)$ to a
constant independent of $\omega$.  This curious phenomenon is shown in
Fig.~\ref{limitzetau}, where we see that $\zeta(z,\omega)$ tends to a constant when
the interaction is not screened.
Inclusion of screening however, will drastically modify this behavior,
reinstating the expected $\omega^2$ behavior (see Fig.~\ref{limitzetas}).
To demonstrate this point we have calculated $\zeta(z,\omega)$ with the bare interaction
$\tilde v(Q,z)$ replaced by a screened interaction according to the
scheme
\begin{equation}\label{screeningscheme}
\tilde v(Q,z) \to \frac{\tilde v(Q,z)}{1-\tilde v(Q) \chi_0(Q,0)}~.
\end{equation}
The denominator is the static dielectric constant of a 2D QW
in which all the electrons reside in the lowest subband, and no
intersubband transitions are allowed.  Indeed  the low-energy
excitations of the 2DEG  are exclusively {\it intra}-subband
electron-hole pairs.  Therefore neglecting {\it inter}-subband
transitions is justified at low frequency and qualitatively correct at
higher frequencies (as long as only a few subbands are involved).

\section{Numerical results}

Fig.~\ref{zetanonlocal} shows the calculated values of the non-local viscosity
$\zeta(z,z',\omega)$ as a function of $z$ and $z'$ for
a sheet density $n_{2D}=10^{11}~\mathrm{cm}^{-2}$  and $L=40$~nm  ($K_F = \sqrt{2 \pi
  n_{2D}}=7.9\times10^5~\mathrm{cm}^{-1}$ ) with  and without the screening (panels a) and b) respectively) .
 We have assumed that the equilibrium density in the QW has the
form   $$n_0(z)=\frac{2}{L}\cos^2\left(\frac{\pi z}{L}\right)~,$$
appropriate to the lowest subband of a free particle in a box.
 This choice gives
\begin{equation}
\tilde v(Q,z)=-\frac{2\pi e^2 L}{\epsilon
  (LQ)^2}\left(\frac{e^{-\frac{QL}{2}}\cosh(Qz)}{1+\frac{(LQ)^2}{4\pi^2}}-\frac{\cos\left(\frac{2\pi
        z}{L}\right)}{1+\frac{4\pi^2}{(QL)^2}}-1\right)
\end{equation}
and, for the unscreened interaction,
\begin{widetext}
\begin{eqnarray}
\tilde v(Q)&=&-\frac{2\pi e^2}{\epsilon Q}\left[\frac{2e^{-\frac{QL}{2}}\sinh(QL/2)}
{(QL)^2\left(1+\frac{(QL)^2}{4\pi^2}\right)^2}-\frac{1}{QL}
-\frac{QL}{16\pi^2\left(1+\frac{(QL)^2}{2\pi^2}\right)}\right].
\end{eqnarray}

The functions $\tilde v(Q,z)$ and $\tilde v(Q)$ depend on the physical
parameter $K_F L$ where $K_F$ is the Fermi momentum of the 2DEG and
$L$ is the thickness of the well in the $z$ direction.  For $K_F L \ll
1 $, $\tilde v(Q,z)\simeq\tilde v(Q)\simeq 2\pi e^2/\epsilon Q$,
i.e. the functions $\tilde v$ reduce to Fourier transforms of the bare
2-dimensional Coulomb  potential.
\begin{center}
\begin{figure}[ht!]
\includegraphics[width=16cm,clip]{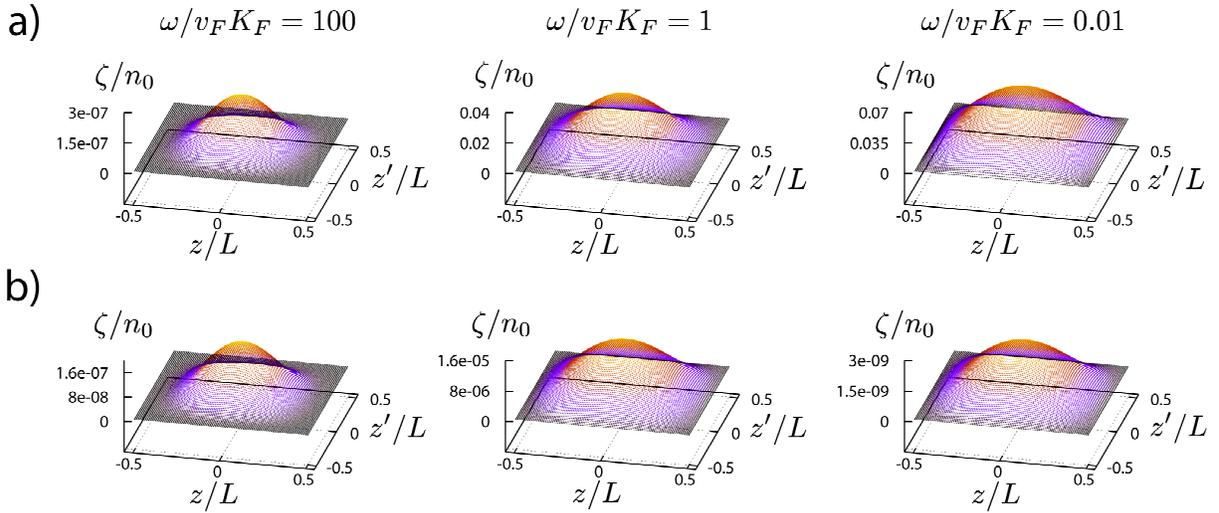}
\caption{The nonlocal viscosity $\zeta(z,z',\omega)/n_0$ as a function of $z,z'$ for
  different values of $\omega$ with $K_F L=3.17$.  Panel   a) -- screened interaction; panel b) unscreened interaction.   Notice the different scales in the two panels.}
\label{zetanonlocal}
\end{figure}
\end{center}
\end{widetext}
The nonlocal viscosity
$\zeta(z,z',\omega)/(n_0(z) \hbar)$  decays when one of the points is
close to the boundary and it is maximum when both  $z$ and $z'$ are
near the center of the well. While the spatial behavior of
$\zeta(z,z',\omega)$ looks qualitatively similar for the screened
[panel a)] and unscreened case [panel b)], the order of magnitude is
quite different, the unscreened results being quite large. This has its origin
in the strong singularity of the Coulomb potential for small momenta.
We then expect that in the case of the unscreened interaction, the dissipation
rate be largely overestimated.

\begin{figure}[ht!] 
\includegraphics[width=8cm,clip]{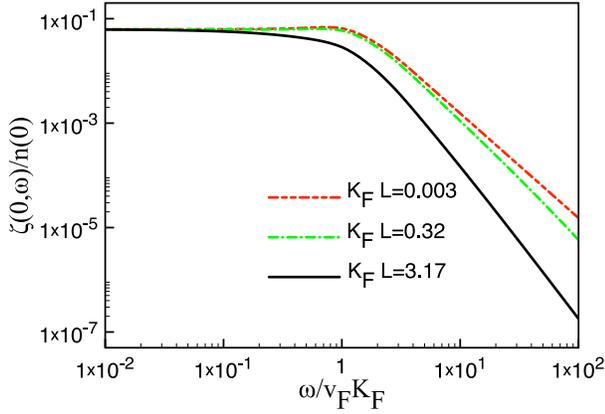}
\caption{Average local viscosity  $\zeta(z,\omega)/\hbar n_0(z)$  from Eq.~(\ref{bulkviscosity-partialaveraged})  as a function of frequency for $z=0$ and for different values of $K_F L$.  The Coulomb interaction potential is unscreened.  Notice that $\zeta$ approaches a finite value for  $\omega \to 0$.}
\label{limitzetau}
\end{figure}

\begin{figure}[ht!]
\includegraphics[width=8cm,clip]{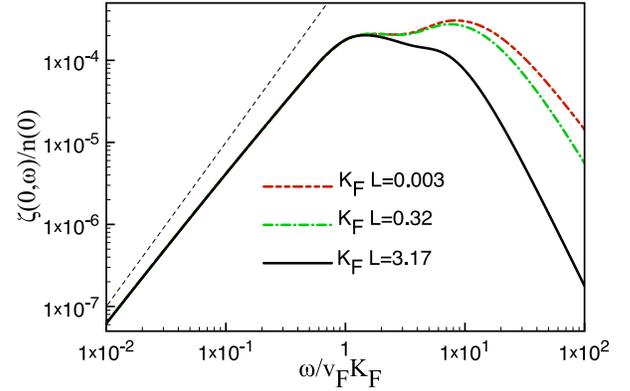}
\caption{Same as Fig.~\ref{limitzetau}, but now for the case of a screened Coulomb interaction. Notice that $\zeta(0,\omega)\propto \omega^2$ for $\omega \to 0$ (the dashed line
is the slope of $\omega^2$ in the logarithmic plot).}
\label{limitzetas}
\end{figure}

In Fig. \ref{limitzetau} we plot the integrated viscosity
$\zeta(0,\omega)$ for various values of  $K_F L$ as a function of the
frequency for the unscreened interaction. For all the values of the
parameter $K_F L$ we notice the existence of a finite limiting value
for small frequencies. As already discussed, this finite limiting
value is due to the existence of a strong singularity for small $Q$,
$1/Q^3$, which compensates for the smallness of the Lindhard
spectra. Including the screening according to the scheme of
Eq.~(\ref{screeningscheme})  changes this behavior dramatically.  The
new behavior is shown in Fig.~\ref{limitzetas} for various values of
the parameter $K_F L$.  Now $\zeta(0,\omega)$ vanishes as $\omega^2$
in agreement with the general discussion of the previous section.

\begin{figure}[ht!]
\includegraphics[width=8cm,clip]{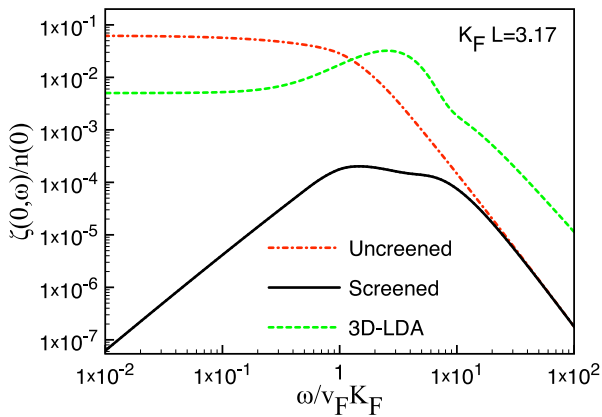}
\caption{Comparison between $\zeta(z,\omega)/\hbar n_0(z)$ and the 3D-LDA viscosity  $Y(z,\omega)$ for $z=0$. As expected the screened and unscreneed results for $\zeta(z,\omega)$
  coincide for large frequencies. Notice that the 3D-LDA (with a screened interaction) predicts a
 finite limit of the viscosity for $\omega\to 0$.}
\label{comparison}
\end{figure}

 Finally in Fig. \ref{comparison} we compare the behavior of
 $\zeta(0,\omega)$ with the corresponding LDA quantity
 $Y(\omega)$,  evaluated at the density $n_0(z=0)$.
In the screened case our viscosity  falls well below $Y(\omega)$
(calculated from the parametrization of Qian and
Vignale\cite{Qian2002})  demonstrating the intrinsic limitation of the
3D-LDA in this regime.

It is also interesting -- although not fully justified physically --
to consider the high-frequency behavior of $\zeta(z,\omega)$ vis-a-vis
the high-frequency behavior of $Y(\omega)$.  This can be
calculated analytically and one sees that the relevant contribution to
the integral over wave vector comes from $Q$ proportional to $\sqrt{2
  m \omega}/\hbar$ and is therefore large. In this regime both $\tilde
v(Q)$ and $v(q)$ scale as $\omega^{-1}$, so the only difference
between the expressions~(\ref{bulkviscosity-partialaveraged}) and~(\ref
{3D-LDAViscosity1}) comes from the different ``volume element"  in
wavevector-space, $Q$ and $q^2$ respectively.  Taking this into
account we immediately understand the origin of the $\sqrt{\omega}$
difference in the high-frequency behaviors of $\zeta(z,\omega)$
($\omega^{-3}$) and $Y(\omega)$   ($\omega^{-5/2}$) --  the
former tending to zero faster than the latter.

To fully evaluate the energy dissipation via either
Eq.~(\ref{energydissipation}) or Eq.~(\ref{energydissipation-LDA})
we need to assign the velocity $v(z,\omega)$. In our model we
consider coherent oscillations of the wave function in the quantum
well. For this motion we then choose
\begin{equation}
\chi(z,t)=\chi_1(z)+\lambda e^{i\omega t}\chi_3(z)
\end{equation}
where $\lambda$ is a constant and $\chi_n(z)$ are the normalized
eigenfunctions of the quantum well
\begin{equation}  
\chi_n(z)=\sqrt{\frac{2}{L}}\left\{
\begin{array}{cc}
\sin\left(\frac{n\pi z}{L}\right) & n~{\mathrm even}\\
\cos\left(\frac{n\pi z}{L}\right) & n~{\mathrm  odd}
\end{array}.
\right.
\end{equation}
For the linear response theory to be valid we have assumed $\lambda\ll 1$.
\footnote{Notice that in the case of the combination $\chi(z)=\chi_1(z)+\lambda e^{i\omega t} \chi_2(z)$
the parity of the state prevents any energy dissipation, i.e., $dE/dt\equiv 0$ for any frequency.}
We can evaluate the current density for the state $\chi(z,t)$ and obtain the velocity field
\begin{equation}
\partial_z v(z,\omega)=\partial_z \frac{j(z,\omega)}{n_0(z)}=-\omega |\lambda|
\cos\left(\frac{2\pi z}{L}\right).
\end{equation}
By substituting this expression in the equation for the energy dissipation, Eq.~(\ref{energydissipation})
we get
\begin{widetext}
\begin{equation}
\frac{dE}{dt}=-\frac{A\hbar}{\pi \omega}\int_0^\omega d\omega' \int dQ~Q
\Im m\chi_0^{2D}(Q,\omega')\Im m\chi_0^{2D}(Q,\omega-\omega')\Gamma^2(Q,\omega')
\label{dissipationrate}
\end{equation}
where we have defined for the unscreened case
\begin{equation}
\Gamma(Q,\omega)=-\frac{e^2\omega}{2\pi \epsilon}\frac{|\lambda|L}{1+\frac{(LQ)^2}{4\pi^2}}
\left[1+\frac{2\pi^2}{(LQ)^2}+\frac{e^{\frac{QL}{2}}\sinh\left(\frac{QL}2\right)}{2QL}
\frac{1-\frac{8\pi^2}{(LQ)^2}}
{\left(1+\frac{(LQ)^2}{16 \pi^2}\right)\left(1+\frac{(LQ)^2}{4\pi^2}\right)}\right].
\end{equation}
\end{widetext}
Accordingly to  the scheme
we adopted [see Eq.~(\ref{screeningscheme})],
to include screening in the interaction potential
we substitute $\Gamma(Q,\omega)$ with
$\Gamma(Q,\omega)/[1-\chi_0(Q)\tilde v(Q)]$.

\begin{figure}[ht!]
\includegraphics[width=8cm,clip]{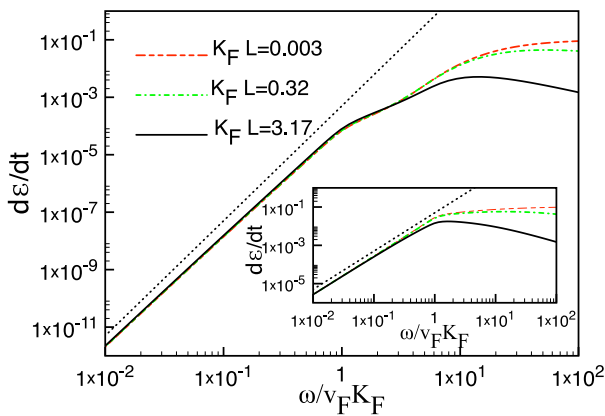}
\caption{The energy dissipation rate for the screened and unscreened
(inset) interaction [see Eq.~(\ref{dissipationrate})]
as a function of the external frequency for various values of the quantity $K_F L$.
The energy dissipation rate scales as $\omega^4$ for the screened interaction, a result
much smaller than the $\omega^2$ behavior obtained from the 3D-LDA energy dissipation rate (also for a screened interaction).
The limiting behavior $d\mathcal{E}/{dt}\simeq\omega^4$ is shown for reference as a dotted line.
Inset: Due to the strong singularity of the unscreened interaction, the energy dissipation rate
for this case, scales as $\omega^2$.  The dotted line shows the limiting
behavior $d\mathcal{E}/{dt}\simeq\omega^2$.}
\label{energyrate}
\end{figure}

In Fig.~\ref{energyrate} we plot $d\mathcal{E}/dt=d/dt (E/ N\hbar \omega_p^2|\lambda|^2)$, where
$\omega_p^2=2\pi n_0(0) e^2 K_F/m$ is the 2D plasmon frequency evaluated at the Fermi momentum,
by directly evaluating Eq.~(\ref{dissipationrate}) for the screened and unscreened case.
It is immediately seen that for small frequencies the energy dissipation rate scales as $\omega^4$ ($\omega^2$)
for the screened (unscreened) interaction. This has to be compared with the energy dissipation rate obtained 
from the 3D-LDA: since $Y(\omega)$ goes to a constant for small frequencies [see Fig.~\ref{comparison}],
we see that the 3D-LDA dissipation rate scales as $\omega^2$ for small $\omega$. Then we can conclude that the 3D-LDA
overestimates the energy dissipation rate in our model.

\section{Discussion}
The system studied in this paper can be thought of as consisting of
two weakly coupled sub-systems of  reduced dimensionality: a linear
oscillator in the $z$ direction coupled to a two-dimensional electron
gas in the $x-y$ plane. We have calculated the energy transfer between
these two sub-systems to leading order in the strength of their
coupling  and we have thus identified a non-local viscosity which can
be compared (after the integration of one variable) with the viscosity
obtained from the standard 3D-LDA.  We have found very significant
differences between the two viscosities, particularly at low
frequency, where the 3D-LDA viscosity is dominated by a shear term which
is absent in the present treatment.

It is important to realize that the problem we have uncovered stems
from the extreme one-dimensional character of the dynamics of our
model.  We have assumed that all the electrons reside in a {\em
single} time-dependent subband $\chi(z,t)$.  In effect, this
dramatically limits the type of excitations we can generate in the
2DEG (see Fig.~\ref{fermisurface}) and this is the reason why we do
not have the transverse current term that is responsible, in three
dimensions, for the shear viscosity term. It is expected --  and
this case is currently under investigation -- that a transversal
shear viscosity term will reappear in a multi-subband system via the
relative motion of the different sheets of the Fermi surface in
different subbands.  For a strictly one-dimensional motion, however,
it seems clear that the use of the LDA is problematic.  This leaves
us with the challenge of formulating a better approximation to keep
track of the systematic reduction in viscosity following from the
geometric confinement of the system.

\acknowledgments R.D'A. thanks C. Ullrich for assistance with part of the 
numerical calculations. R.D'A. and M. Di V. acknowledge financial support by the Department of
Energy grant DE-FG02-05ER46204. G. V. acknowledges financial support by the Department of Energy grant DE-FG02-05ER46203.

\bibliography{mine,articles,books}

\begin{thebibliography}{26}
\expandafter\ifx\csname natexlab\endcsname\relax\def\natexlab#1{#1}\fi
\expandafter\ifx\csname bibnamefont\endcsname\relax
  \def\bibnamefont#1{#1}\fi
\expandafter\ifx\csname bibfnamefont\endcsname\relax
  \def\bibfnamefont#1{#1}\fi
\expandafter\ifx\csname citenamefont\endcsname\relax
  \def\citenamefont#1{#1}\fi
\expandafter\ifx\csname url\endcsname\relax
  \def\url#1{\texttt{#1}}\fi
\expandafter\ifx\csname urlprefix\endcsname\relax\def\urlprefix{URL }\fi
\providecommand{\bibinfo}[2]{#2}
\providecommand{\eprint}[2][]{\url{#2}}

\bibitem[{\citenamefont{Runge and Gross}(1984)}]{Runge1984}
\bibinfo{author}{\bibfnamefont{E.}~\bibnamefont{Runge}} \bibnamefont{and}
  \bibinfo{author}{\bibfnamefont{E.~K.~U.} \bibnamefont{Gross}},
  \bibinfo{journal}{Phys. Rev. Lett.} \textbf{\bibinfo{volume}{52}},
  \bibinfo{pages}{997} (\bibinfo{year}{1984}).

\bibitem[{\citenamefont{Marques et~al.}(2006)\citenamefont{Marques, Ullrich,
  Nogueira, Rubio, Burke, and Gross}}]{Marques}
\bibinfo{editor}{\bibfnamefont{M.}~\bibnamefont{Marques}},
  \bibinfo{editor}{\bibfnamefont{C.}~\bibnamefont{Ullrich}},
  \bibinfo{editor}{\bibfnamefont{F.}~\bibnamefont{Nogueira}},
  \bibinfo{editor}{\bibfnamefont{A.}~\bibnamefont{Rubio}},
  \bibinfo{editor}{\bibfnamefont{K.}~\bibnamefont{Burke}}, \bibnamefont{and}
  \bibinfo{editor}{\bibfnamefont{E.~K.~U.} \bibnamefont{Gross}}, eds.,
  \emph{\bibinfo{title}{Time-Dependent Density Functional Theory}}, vol.
  \bibinfo{volume}{706/2006} of \emph{\bibinfo{series}{Lecture Notes in
  Physics}} (\bibinfo{publisher}{Springer Berlin / Heidelberg},
  \bibinfo{year}{2006}).

\bibitem[{\citenamefont{Giuliani and Vignale}(2005)}]{Giulianivignale}
\bibinfo{author}{\bibfnamefont{G.~F.} \bibnamefont{Giuliani}} \bibnamefont{and}
  \bibinfo{author}{\bibfnamefont{G.}~\bibnamefont{Vignale}},
  \emph{\bibinfo{title}{Quantum Theory of the Electron Liquid}}
  (\bibinfo{publisher}{Cambridge}, \bibinfo{address}{Cambridge, UK},
  \bibinfo{year}{2005}).

\bibitem[{\citenamefont{Hohenberg and Kohn}(1964)}]{Hohenberg1964}
\bibinfo{author}{\bibfnamefont{P.}~\bibnamefont{Hohenberg}} \bibnamefont{and}
  \bibinfo{author}{\bibfnamefont{W.}~\bibnamefont{Kohn}},
  \bibinfo{journal}{Phys. Rev.} \textbf{\bibinfo{volume}{136}},
  \bibinfo{pages}{B864} (\bibinfo{year}{1964}).

\bibitem[{\citenamefont{Kohn and Sham}(1965)}]{Kohn1965}
\bibinfo{author}{\bibfnamefont{W.}~\bibnamefont{Kohn}} \bibnamefont{and}
  \bibinfo{author}{\bibfnamefont{L.~J.} \bibnamefont{Sham}},
  \bibinfo{journal}{Phys. Rev.} \textbf{\bibinfo{volume}{140}},
  \bibinfo{pages}{A1133} (\bibinfo{year}{1965}).

\bibitem[{\citenamefont{van Leeuwen}(1999)}]{vanLeeuwen1999}
\bibinfo{author}{\bibfnamefont{R.}~\bibnamefont{van Leeuwen}},
  \bibinfo{journal}{Phys. Rev. Lett.} \textbf{\bibinfo{volume}{82}},
  \bibinfo{pages}{3863} (\bibinfo{year}{1999}).

\bibitem[{\citenamefont{Vignale and Kohn}(1996)}]{Vignale1996}
\bibinfo{author}{\bibfnamefont{G.}~\bibnamefont{Vignale}} \bibnamefont{and}
  \bibinfo{author}{\bibfnamefont{W.}~\bibnamefont{Kohn}}, in
  \emph{\bibinfo{booktitle}{Electronic density functional Theory: recent
  progress and new directions}}, edited by
  \bibinfo{editor}{\bibfnamefont{J.~F.} \bibnamefont{Dobson}},
  \bibinfo{editor}{\bibfnamefont{G.}~\bibnamefont{Vignale}}, \bibnamefont{and}
  \bibinfo{editor}{\bibfnamefont{M.~P.} \bibnamefont{Das}}
  (\bibinfo{publisher}{Plenum}, \bibinfo{address}{N.Y.}, \bibinfo{year}{1996}),
  p. \bibinfo{pages}{199}.

\bibitem[{\citenamefont{Ghosh and Dhara}(1988)}]{Ghosh1988}
\bibinfo{author}{\bibfnamefont{S.~K.} \bibnamefont{Ghosh}} \bibnamefont{and}
  \bibinfo{author}{\bibfnamefont{A.~K.} \bibnamefont{Dhara}},
  \bibinfo{journal}{\pra} \textbf{\bibinfo{volume}{38}}, \bibinfo{pages}{1149}
  (\bibinfo{year}{1988}).

\bibitem[{\citenamefont{Vignale et~al.}(1997)\citenamefont{Vignale, Ullrich,
  and Conti}}]{Vignale1997b}
\bibinfo{author}{\bibfnamefont{G.}~\bibnamefont{Vignale}},
  \bibinfo{author}{\bibfnamefont{C.~A.} \bibnamefont{Ullrich}},
  \bibnamefont{and} \bibinfo{author}{\bibfnamefont{S.}~\bibnamefont{Conti}},
  \bibinfo{journal}{Phys. Rev. Lett.} \textbf{\bibinfo{volume}{79}},
  \bibinfo{pages}{4878} (\bibinfo{year}{1997}).

\bibitem[{\citenamefont{D'Agosta and Di~Ventra}(2006)}]{DAgosta2006a}
\bibinfo{author}{\bibfnamefont{R.}~\bibnamefont{D'Agosta}} \bibnamefont{and}
  \bibinfo{author}{\bibfnamefont{M.}~\bibnamefont{Di~Ventra}},
  \bibinfo{journal}{J. Phys.: Cond. Matt.} \textbf{\bibinfo{volume}{18}},
  \bibinfo{pages}{11059} (\bibinfo{year}{2006}).

\bibitem[{\citenamefont{Williams et~al.}(2001)\citenamefont{Williams, Sherwin,
  Maranowski, and Gossard}}]{Williams2001}
\bibinfo{author}{\bibfnamefont{J.~B.} \bibnamefont{Williams}},
  \bibinfo{author}{\bibfnamefont{M.~S.} \bibnamefont{Sherwin}},
  \bibinfo{author}{\bibfnamefont{K.~D.} \bibnamefont{Maranowski}},
  \bibnamefont{and} \bibinfo{author}{\bibfnamefont{A.~C.}
  \bibnamefont{Gossard}}, \bibinfo{journal}{Phys. Rev. Lett.}
  \textbf{\bibinfo{volume}{87}}, \bibinfo{pages}{037401}
  (\bibinfo{year}{2001}).

\bibitem[{\citenamefont{Ullrich and Vignale}(2001)}]{Ullrich2001}
\bibinfo{author}{\bibfnamefont{C.~A.} \bibnamefont{Ullrich}} \bibnamefont{and}
  \bibinfo{author}{\bibfnamefont{G.}~\bibnamefont{Vignale}},
  \bibinfo{journal}{Phys. Rev. Lett.} \textbf{\bibinfo{volume}{87}},
  \bibinfo{pages}{037402} (\bibinfo{year}{2001}).

\bibitem[{\citenamefont{Ullrich and Vignale}(2002)}]{Ullrich2002}
\bibinfo{author}{\bibfnamefont{C.~A.} \bibnamefont{Ullrich}} \bibnamefont{and}
  \bibinfo{author}{\bibfnamefont{G.}~\bibnamefont{Vignale}},
  \bibinfo{journal}{\prb} \textbf{\bibinfo{volume}{65}},
  \bibinfo{pages}{245102} (\bibinfo{year}{2002}).

\bibitem[{\citenamefont{Ullrich and Burke}(2004)}]{Ullrich2004}
\bibinfo{author}{\bibfnamefont{C.~A.} \bibnamefont{Ullrich}} \bibnamefont{and}
  \bibinfo{author}{\bibfnamefont{K.}~\bibnamefont{Burke}}, \bibinfo{journal}{J.
  Chem. Phys.} \textbf{\bibinfo{volume}{121}}, \bibinfo{pages}{28}
  (\bibinfo{year}{2004}).

\bibitem[{\citenamefont{van Faassen et~al.}(2002)\citenamefont{van Faassen,
  de~Boeij, van Leeuwen, Berger, and Snijders}}]{vanFaassen2002}
\bibinfo{author}{\bibfnamefont{M.}~\bibnamefont{van Faassen}},
  \bibinfo{author}{\bibfnamefont{P.~L.} \bibnamefont{de~Boeij}},
  \bibinfo{author}{\bibfnamefont{R.}~\bibnamefont{van Leeuwen}},
  \bibinfo{author}{\bibfnamefont{J.~A.} \bibnamefont{Berger}},
  \bibnamefont{and} \bibinfo{author}{\bibfnamefont{J.~G.}
  \bibnamefont{Snijders}}, \bibinfo{journal}{Phys. Rev. Lett.}
  \textbf{\bibinfo{volume}{88}}, \bibinfo{pages}{186401}
  (\bibinfo{year}{2002}).

\bibitem[{\citenamefont{Berger et~al.}(2005)\citenamefont{Berger, de~Boeij, and
  van Leeuwen}}]{Berger2005}
\bibinfo{author}{\bibfnamefont{J.~A.} \bibnamefont{Berger}},
  \bibinfo{author}{\bibfnamefont{P.~L.} \bibnamefont{de~Boeij}},
  \bibnamefont{and} \bibinfo{author}{\bibfnamefont{R.}~\bibnamefont{van
  Leeuwen}}, \bibinfo{journal}{Phys. Rev. B} \textbf{\bibinfo{volume}{71}},
  \bibinfo{pages}{155104} (\bibinfo{year}{2005}).

\bibitem[{\citenamefont{Berger et~al.}(2007)\citenamefont{Berger, de~Boeij, and
  van Leeuwen}}]{Berger2007}
\bibinfo{author}{\bibfnamefont{J.~A.} \bibnamefont{Berger}},
  \bibinfo{author}{\bibfnamefont{P.~L.} \bibnamefont{de~Boeij}},
  \bibnamefont{and} \bibinfo{author}{\bibfnamefont{R.}~\bibnamefont{van
  Leeuwen}}, \bibinfo{journal}{Phys. Rev. B} \textbf{\bibinfo{volume}{75}},
  \bibinfo{pages}{035116} (\bibinfo{year}{2007}).

\bibitem[{\citenamefont{Sai et~al.}(2005)\citenamefont{Sai, Zwolak, Vignale,
  and Di~Ventra}}]{Sai2005}
\bibinfo{author}{\bibfnamefont{N.}~\bibnamefont{Sai}},
  \bibinfo{author}{\bibfnamefont{M.}~\bibnamefont{Zwolak}},
  \bibinfo{author}{\bibfnamefont{G.}~\bibnamefont{Vignale}}, \bibnamefont{and}
  \bibinfo{author}{\bibfnamefont{M.}~\bibnamefont{Di~Ventra}},
  \bibinfo{journal}{Phys. Rev. Lett.} \textbf{\bibinfo{volume}{94}},
  \bibinfo{pages}{186810} (\bibinfo{year}{2005}).

\bibitem[{\citenamefont{D'Agosta et~al.}(2006)\citenamefont{D'Agosta, Sai, and
  Di~Ventra}}]{DAgosta2006c}
\bibinfo{author}{\bibfnamefont{R.}~\bibnamefont{D'Agosta}},
  \bibinfo{author}{\bibfnamefont{N.}~\bibnamefont{Sai}}, \bibnamefont{and}
  \bibinfo{author}{\bibfnamefont{M.}~\bibnamefont{Di~Ventra}},
  \bibinfo{journal}{Nano Lett.} \textbf{\bibinfo{volume}{6}},
  \bibinfo{pages}{2935} (\bibinfo{year}{2006}).

\bibitem[{\citenamefont{Sai et~al.}(2007)\citenamefont{Sai, Bushong, Hatcher,
  and Di~Ventra}}]{Sai2007}
\bibinfo{author}{\bibfnamefont{N.}~\bibnamefont{Sai}},
  \bibinfo{author}{\bibfnamefont{N.}~\bibnamefont{Bushong}},
  \bibinfo{author}{\bibfnamefont{R.}~\bibnamefont{Hatcher}}, \bibnamefont{and}
  \bibinfo{author}{\bibfnamefont{M.}~\bibnamefont{Di~Ventra}},
  \bibinfo{journal}{cond-mat/0701634}  (\bibinfo{year}{2007}).

\bibitem[{\citenamefont{Romaniello and de~Boeij}(2005)}]{Romaniello2005}
\bibinfo{author}{\bibfnamefont{P.}~\bibnamefont{Romaniello}} \bibnamefont{and}
  \bibinfo{author}{\bibfnamefont{P.~L.} \bibnamefont{de~Boeij}},
  \bibinfo{journal}{Phys. Rev. B} \textbf{\bibinfo{volume}{71}},
  \bibinfo{pages}{155108} (\bibinfo{year}{2005}).

\bibitem[{\citenamefont{Berger et~al.}(2006)\citenamefont{Berger, Romaniello,
  van Leeuwen, and de~Boeij}}]{Berger2006}
\bibinfo{author}{\bibfnamefont{J.~A.} \bibnamefont{Berger}},
  \bibinfo{author}{\bibfnamefont{P.}~\bibnamefont{Romaniello}},
  \bibinfo{author}{\bibfnamefont{R.}~\bibnamefont{van Leeuwen}},
  \bibnamefont{and} \bibinfo{author}{\bibfnamefont{P.~L.}
  \bibnamefont{de~Boeij}}, \bibinfo{journal}{Phys. Rev. B}
  \textbf{\bibinfo{volume}{74}}, \bibinfo{pages}{245117}
  (\bibinfo{year}{2006}).

\bibitem[{\citenamefont{D'Agosta and Vignale}(2006)}]{DAgosta2006}
\bibinfo{author}{\bibfnamefont{R.}~\bibnamefont{D'Agosta}} \bibnamefont{and}
  \bibinfo{author}{\bibfnamefont{G.}~\bibnamefont{Vignale}},
  \bibinfo{journal}{Phys. Rev. Lett.} \textbf{\bibinfo{volume}{96}},
  \bibinfo{pages}{016405} (\bibinfo{year}{2006}).

\bibitem[{\citenamefont{Wijewardane and Ullrich}(2005)}]{Wijewardane2005}
\bibinfo{author}{\bibfnamefont{H.~O.} \bibnamefont{Wijewardane}}
  \bibnamefont{and} \bibinfo{author}{\bibfnamefont{C.~A.}
  \bibnamefont{Ullrich}}, \bibinfo{journal}{Phys. Rev. Lett.}
  \textbf{\bibinfo{volume}{95}}, \bibinfo{pages}{086401}
  (\bibinfo{year}{2005}).

\bibitem[{\citenamefont{Conti and Vignale}(1999)}]{Conti1999}
\bibinfo{author}{\bibfnamefont{S.}~\bibnamefont{Conti}} \bibnamefont{and}
  \bibinfo{author}{\bibfnamefont{G.}~\bibnamefont{Vignale}},
  \bibinfo{journal}{Phys. Rev. B} \textbf{\bibinfo{volume}{60}},
  \bibinfo{pages}{7966} (\bibinfo{year}{1999}).

\bibitem[{\citenamefont{Qian and Vignale}(2002)}]{Qian2002}
\bibinfo{author}{\bibfnamefont{Z.}~\bibnamefont{Qian}} \bibnamefont{and}
  \bibinfo{author}{\bibfnamefont{G.}~\bibnamefont{Vignale}},
  \bibinfo{journal}{Phys. Rev. B} \textbf{\bibinfo{volume}{65}},
  \bibinfo{pages}{235121} (\bibinfo{year}{2002}).

\end{thebibliography}
\end{document}